\documentclass[12pt]{nature}

\usepackage{times}

\usepackage{caption}

\usepackage{graphicx}

\title{Studying ultrafast Rabi dynamics with a short-wavelength seeded free-electron laser}

\author{Saikat Nandi$^{1\ast}$, Edvin Olofsson$^2$, Mattias Bertolino$^2$, Stefanos Carlstr\"{o}m$^{2}$, Felipe Zapata$^{2}$, David Busto$^{2}$, Carlo Callegari$^{3}$, Michele Di Fraia$^{3}$, Per Eng-Johnsson$^{2}$, Raimund Feifel$^{4}$, Guillaume Gallician$^{5}$,  Mathieu Gisselbrecht$^{2}$, Sylvain Maclot$^{2,4}$, Lana Neori\v{c}i\'{c}$^{2}$, Jasper Peschel$^{2}$, Oksana Plekan$^{3}$, Kevin C. Prince$^{3}$, Richard J. Squibb$^{4}$, Shiyang Zhong$^{2}$, Philipp V. Demekhin$^{6}$, Michael Meyer$^{7}$, Catalin Miron$^{5,8}$, Laura Badano$^{3}$, Miltcho B. Danailov$^{3}$, Luca Giannessi$^{3,9}$, Michele Manfredda,$^{3}$ Filippo Sottocorona$^{3,10}$, Marco Zangrando,$^{3,11}$ \& Jan Marcus Dahlstr\"{o}m$^{2\ast}$}

\begin{document}

\maketitle

\begin{affiliations}
 \item Universit\'{e} de Lyon, Universit\'{e} Claude Bernard Lyon 1, CNRS, Institut Lumi\`{e}re Mati\`{e}re, Villeurbanne, France
 \item Department of Physics, Lund University, Lund, Sweden.
 \item Elettra-Sincrotrone Trieste, Basovizza, Trieste, Italy
 \item Department of Physics, University of Gothenburg, Gothenburg, Sweden
 \item Universit\'{e} Paris-Saclay, CEA, CNRS, LIDYL, Gif-sur-Yvette, France
 \item Institute of Physics and CINSaT, University of Kassel, Heinrich-Plett-Stra{\ss}e 40, Kassel, Germany
 \item European XFEL, Schenefeld, Germany
 \item ELI-NP, ``Horia Hulubei'' National Institute for Physics and Nuclear Engineering, M\v{a}gurele, Romania
 \item Istituto Nazionale di Fisica Nucleare, Laboratori Nazionali di Frascati, Frascati, Italy
 \item Universit\`{a} degli Studi di Trieste, Trieste, Italy
 \item IOM-CNR, Istituto Officina dei Materiali, Basovizza, Trieste, Italy
\end{affiliations}

\begin{abstract}
Rabi oscillations are periodic modulations of populations in two-level systems interacting with a time-varying field\cite{rabi1937}. They are ubiquitous in physics with applications in different areas such as photonics\cite{fushman2008}, nano-electronics\cite{vijay2012}, electron microscopy\cite{feist2015}, and quantum information\cite{pedrozo2020}. While the theory developed by Rabi was intended for fermions in gyrating magnetic fields, Autler and Townes realized that it could also be used to describe coherent light--matter interaction within the rotating wave approximation\cite{autler1955}. Although intense nanometer-wavelength light-sources have been available for more than a decade\cite{ackermann2007,emma2010,ishikawa2012}, Rabi dynamics at such short wavelengths have not been observed directly. Here we show that femtosecond extreme-ultraviolet pulses from a seeded free-electron laser\cite{allaria2012} can drive Rabi oscillations between the ground state and an excited state in helium atoms. The measured photoemission signal revealed an Autler-Townes doublet as well as an avoided crossing, phenomena that are both trademarks of quantum optics\cite{haroche2006}. Using theoretical analyses that go beyond the strong-field approximation\cite{reiss1980}, we found that the ultrafast build-up of the doublet structure follows from a quantum interference effect between resonant and non-resonant photoionization pathways. Given the recent availability of intense attosecond\cite{maroju2020} and few-femtosecond\cite{mirian2021} extreme-ultraviolet pulses, our results offer opportunities to carry out ultrafast manipulation of coherent processes at short wavelengths using free-electron lasers.
\end{abstract}

The advent of free-electron laser (FEL) facilities, providing femtosecond light pulses in the gigawatt regime at extreme-ultraviolet (XUV) or X-ray wavelengths, has opened up new prospects for experiments in isolated atoms and molecules in the gas-phase\cite{young2018,lindroth2019}. Over the last decade, pioneering results concerning multi-photon ionization of atoms\cite{young2010}, and small molecules\cite{rudenko2017} were obtained using pulses from self-amplified spontaneous emission (SASE) FEL sources\cite{emma2010}. However, these pulses are prone to low degree of coherence and poor shot-to-shot reproducibility due to the inherent instability pertinent to the SASE process. As a result, despite theoretical predictions to observe Rabi oscillations at short wavelengths\cite{lagattuta1993,girju2007,rohringer2008,zhang2014}, effects from these oscillations on the measured spectra were only indirect\cite{sako2011,kanter2011}.
Instead, XUV pulses from a SASE FEL have been used as pumps that allowed subsequent ultrafast Rabi dynamics to be driven by laser pulses at near-infrared wavelengths\cite{fushitani2016}.
In this regard, XUV pulses from a seeded FEL, such as FERMI\cite{allaria2012}, with its high temporal and spatial coherence, and large peak intensity can permit studying coherent light--matter interactions\cite{prince2016} as well as phase-dependent interference effects of the wavefunction\cite{fushman2008}.  

According to the Rabi model\cite{rabi1937}, if a two-level atom initially in its ground state $|a\rangle$, is subjected to an interaction with a field of frequency $\omega$ that couples it to the excited state $|b\rangle$, the probability for excitation varies sinusoidally in time as: $P_b(t)=|\frac{\Omega}{W}|^2\sin^2\left(\frac{Wt}{2}\right)$. 
The oscillating population leads to a symmetric doublet structure in the frequency domain, known as an Autler-Townes (AT) doublet. The splitting is given by the generalized Rabi frequency: $W = \sqrt{\Omega^2+\Delta\omega^2}$, where $\Delta\omega=\omega-\omega_{ba}$, is the detuning of the field with respect to the atomic transition frequency, $\omega_{ba}$. The Rabi frequency for light-matter interaction within the dipole approximation is: $\Omega=eE_0 z_{ba}/\hbar$,  with $E_0$ being the electric field amplitude, $z_{ba}$ the transition matrix element, $\hbar$ the reduced Planck's constant and $e$ the elementary charge. 
In addition to the periodic population transfer $P_b(t)$, the coherent dynamics is further associated with {\it sign changes} of the oscillating amplitudes for the two states. For fermions, such sign changes of the wavefunction can be connected to rotations in real space\cite{rabi1937} that have been measured for neutron beams in magnetic fields\cite{rauch1975}. Analogous sign changes in quantum optics were studied using Rydberg atoms to determine the number of photons in a cavity\cite{nogues1999}. Recently, the sign changes in Rabi amplitudes have been predicted to strongly alter AT doublet structures in photo-excited atoms, when probed by attosecond XUV pulses\cite{jiang2021}. 

\begin{figure*}[!htb]
\centering
\includegraphics[width=\textwidth]{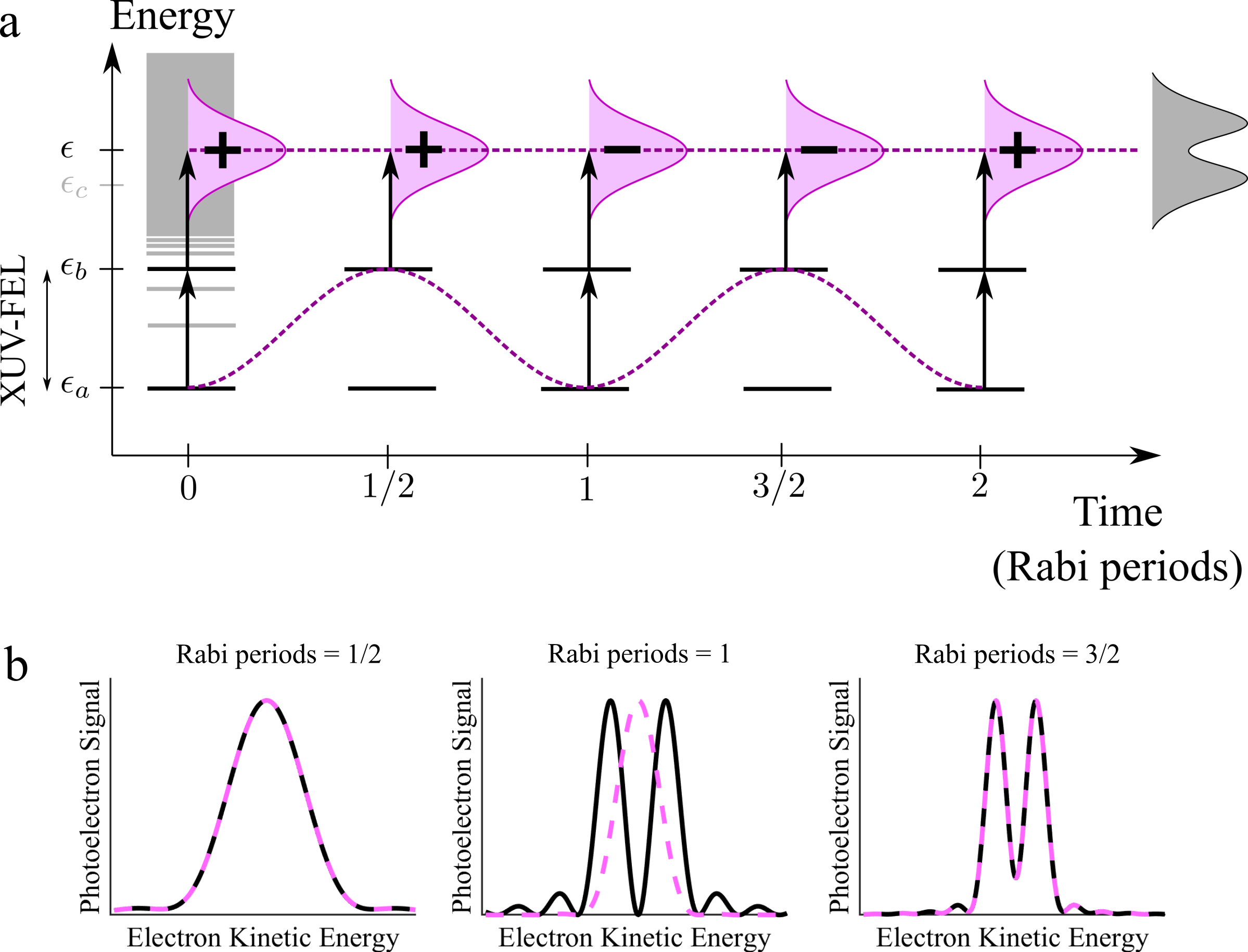}
\caption{{\bf Rabi oscillations induced by an XUV-FEL pulse.} {\bf a},  The sinusoidal population transfer between the XUV-FEL coupled states: $|a\rangle$ and $|b\rangle$ (black horizontal lines) is associated with sign changes between adjacent Rabi cycles.  Photoelectrons can be ejected from excited state $|b\rangle$, by one photon, or by two photons from $|a\rangle$ via intermediate states, $|c\rangle$ (grey horizontal lines). This results in the build up of an ultrafast Autler-Townes doublet structure. {\bf b}, The build up of an ultrafast Autler-Townes doublet for $1/2$, $1$ and $3/2$ completed Rabi periods is shown for one-photon ionization from $|b\rangle$ (dashed, magenta line) and two-photon ionization from $|a\rangle$ (solid, black line) using the analytical model described in the Supplementary Information.}  
\label{fig1}
\end{figure*}

Here, we investigate the Rabi dynamics at XUV wavelengths in helium atoms induced by an intense pulse from the FERMI seeded FEL that couples the two levels: $|a\rangle=1s^2$ ($^1S_0$) and $|b\rangle=1s4p$ ($^1P_1$), with $\hbar\omega_{ba}=\epsilon_b-\epsilon_a= 23.742$ eV \cite{kramida2021}. The dynamics is probed in-situ by recording photoelectrons ejected from the state $|b\rangle$ or $|a\rangle$ during the ultrashort interaction, with one or two XUV-FEL photons, as illustrated in Fig.~\ref{fig1}a.
In order to interpret this non-linear dynamics, we have developed an analytical model based on a Dyson series for the two-level system undergoing Rabi oscillations (see Supplementary Information for details). The resulting AT doublet structure depends on whether the photoelectron is originating from the ground state, $|a\rangle$, or the excited state, $|b\rangle$, as shown in Fig.~\ref{fig1}b. 
The narrow spectral bandwidth of the XUV-FEL pulse ($20$ - $65$ meV; see Methods) enables efficient coupling of $|a\rangle$ and $|b\rangle$ with the dipole element:  $z_{ba}=0.1318 a_0$, $a_0$ being the Bohr radius\cite{chan1991}. It is possible to drive the transition {\it coherently}, because the excited state lifetime ($\sim 4$ ns) is much longer than the estimated full-width-at-half-maximum (FWHM) of the FEL pulse duration\cite{finetti2017}: $56\pm 13$ fs. 
Thus, the Hamiltonian for a two-level system: $H=\frac{1}{2}\hbar\omega_{ba} \widehat{\sigma}_z+\hbar\Omega\sin(\omega t)\widehat{\sigma}_x$, where $\widehat{\sigma}_z$ and $\widehat{\sigma}_x$ are Pauli operators, can be satisfied by a time-dependent wavefunction of the form: $|\Psi(t)\rangle=a(t)e^{-i\epsilon_a t/\hbar}|a\rangle+b(t)e^{-i\epsilon_b t/\hbar}|b\rangle$. Within the rotating wave approximation (RWA), the amplitudes of the ground and excited states are expressed as:
\begin{equation}
\left\{\begin{array}{ll}
a(t)=& \left[\cos\frac{W t}{2}-i\frac{\Delta\omega}{W}\sin\frac{Wt}{2}\right]\exp(i\Delta \omega t /2) \\
b(t)=& -i\frac{\Omega}{W}\sin\frac{W t}{2}\exp(-i \Delta\omega t/2),
\end{array}\right.
\label{eq:rabi}
\end{equation}
provided that the electric field can be approximated as a flat-top shape in time.
The sign changes associated with these Rabi amplitudes are essential to understand the ultrafast build-up of AT doublets from $|a\rangle$ or $|b\rangle$, by absorption of two or one resonant XUV-FEL photons. 
The AT doublet emerges due to a destructive interference effect between photoelectrons ejected before and after the first sign change, which is found to occur at $1/2$ and $1$ Rabi period for the amplitudes: $a(t)$ and $b(t)$ with $\Delta \omega=0$, respectively. This is in agreement with the results from the analytical model presented in Fig.~\ref{fig1}b.

\begin{figure*}[!htb]
\centering
\includegraphics[width=\textwidth]{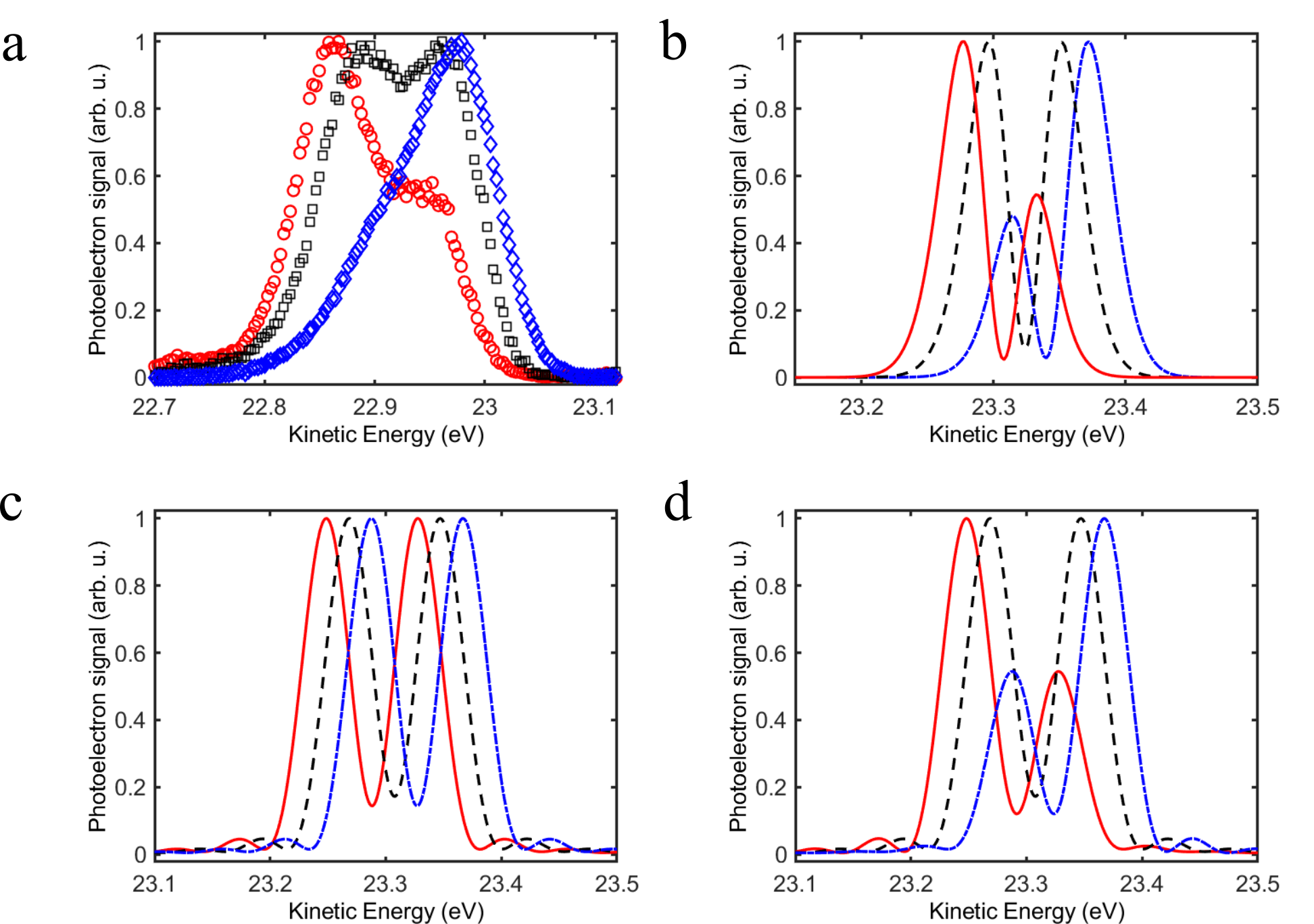}
\caption{{\bf Asymmetry of the ultrafast Autler-Townes doublet.} {\bf a}, Deconvoluted experimental photoelectron spectra with symmetric AT doublet (black squares) at $23.753$ eV photon energy, and the asymmetric ones at $\pm 13$ meV detuning (blue diamonds and red circles, respectively). {\bf b}, Ab initio photoelectron spectra using TDCIS at three photon energies with symmetric AT doublet at $24.157$ eV (dashed, black line) and the asymmetric ones at $\pm 13$ meV detuning. Red (blue) curve corresponds to red (blue) detuned light. {\bf c}-{\bf d}, Same as {\bf b}, but using the analytical model for $3/2$ Rabi periods in case of one-photon ionization from $|b\rangle$ and two-photon ionization from $|a\rangle$, respectively.}   
\label{fig2}
\end{figure*}

Measured photoelectron spectra, displayed in Fig.~\ref{fig2}a, exhibit an AT splitting of $\hbar\Omega = 80\pm 2$~meV (see Methods for details about the blind deconvolution procedure used here). The corresponding Rabi period: $2\pi/\Omega\approx 52$ fs, given its proximity to the FWHM of the XUV-FEL pulse, suggests that the experiment was performed in a regime of ultrafast AT doublet formation close to a {\it single} Rabi cycle. A slight blue detuning of the XUV light by almost $11$ meV, relative to the atomic transition, is required to record a symmetric AT doublet (black squares in Fig.~\ref{fig2}a). A strong asymmetry is observed when the FEL frequency is detuned to the red (red circles in Fig.~\ref{fig2}a) or blue (blue diamonds in Fig.~\ref{fig2}a) side of the symmetric doublet. 
The asymmetry of the AT doublet is quantitatively well-reproduced by ab initio numerical simulations for helium within the time-dependent configuration-interaction singles (TDCIS) approximation\cite{greenman2010}, as shown in Fig.~\ref{fig2}b. Gaussian pulses were used with parameters chosen to match the experimental conditions with effective intensity of $2\times10^{13}$ W/cm$^2$ (as obtained from $\Omega$) and pulse duration (FWHM) $=56$ fs (see Methods for details). 
It is worth noting that the Rabi dynamics is sensitive to the exact shape of the driving pulse. For instance, a Gaussian pulse can induce more Rabi oscillations than a flat-top pulse with same FWHM by a factor of $\sqrt{\pi/(2\ln 2)}\approx 1.5$. 
Thus, the calculated photoelectron spectra from the analytical model using flat-top pulses in Fig.~\ref{fig1}b for $3/2$ Rabi periods agree well with those from the TDCIS calculations using Gaussian pulses having FWHM close to a single Rabi period in Fig.~\ref{fig2}b. Clearly, the AT doublet manifests itself between $1$ and $3/2$ Rabi periods.
The difference in kinetic energy ($\sim 0.4$ eV) of the symmetric AT doublet between experiment and theory (see Fig.~\ref{fig2}a and 2b, respectively) is attributed to electron correlation effects not included in the TDCIS calculations that increase the binding energy beyond the Hartree-Fock level.  
The observed asymmetry in the AT doublet cannot be explained by a breakdown of the RWA because the experiment is performed at a resonant weak-coupling condition\cite{autler1955}: $\omega\approx\omega_{ba}$ and $\Omega/\omega_{ba}=0.34\%$.   
Instead, we express the Rabi amplitudes from Eq.~\ref{eq:rabi} in terms of their frequency components and find that $a(t)$ has two asymmetric components that are proportional to $(1\pm\Delta\omega/W)$, while $b(t)$ has two symmetric components that are proportional to $\pm\Omega/W$. 
Using the analytical model with $3/2$ Rabi periods, we confirm that the AT doublet from $|b\rangle$ is symmetric, while that from $|a\rangle$ is asymmetric, as shown in Fig.~\ref{fig2}c and 2d, respectively. 
Quite remarkably, the observed asymmetry in the experiment suggests that the photoelectron signal contains significant contributions from the two-photon ionization process from $|a\rangle$. How is this possible given that the electric field amplitude: $E_0=0.02388$ a.u. implies an ionization-probability ratio of $10^4:1$ in favor of the one-photon process from $|b\rangle$?

\begin{figure*}[!htb]
\centering
\includegraphics[width=\textwidth]{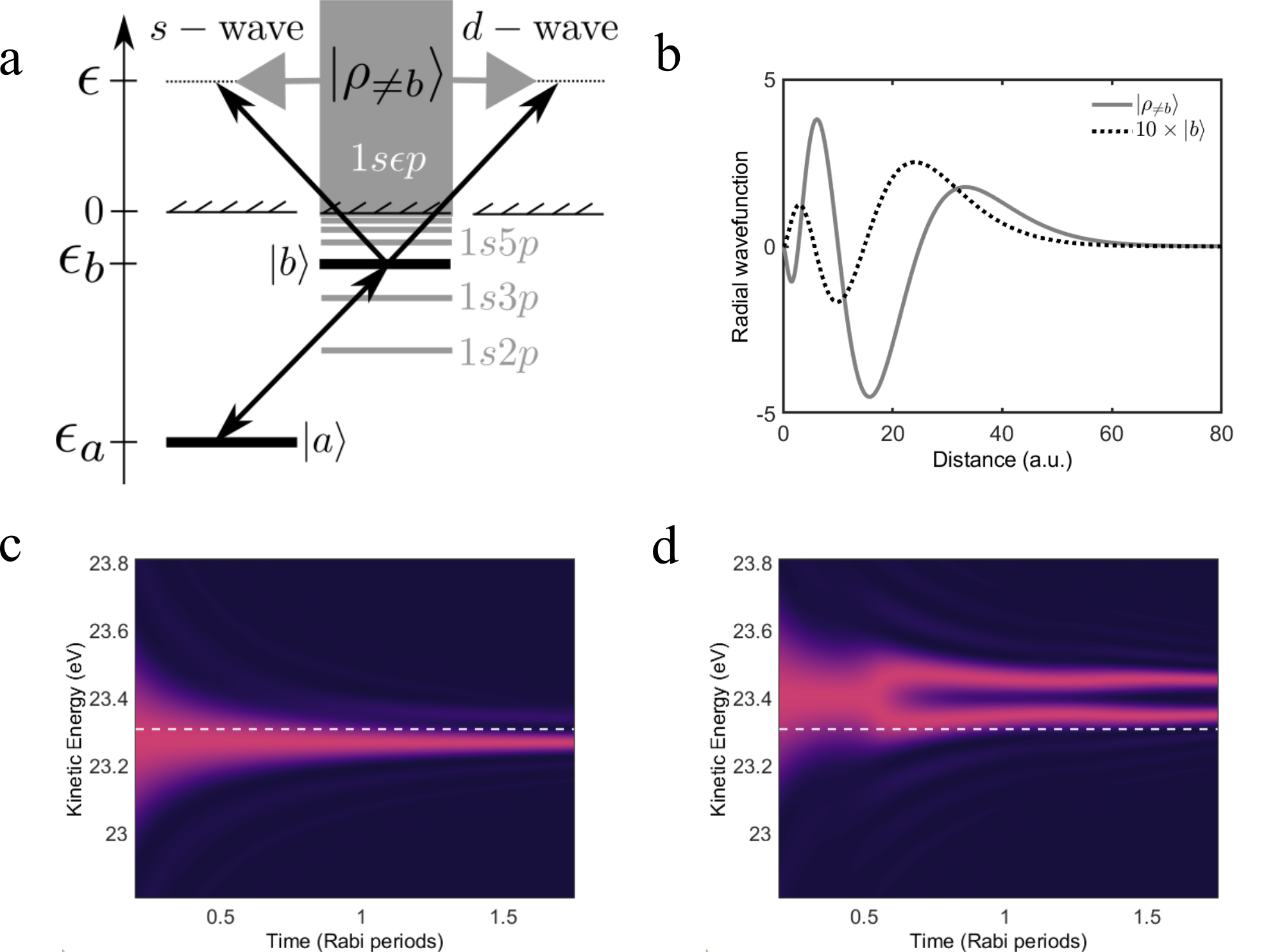}
\caption{{\bf Quantum interference with a giant wave. }{\bf a}, Energy level diagram for the photon transitions that lead to quantum interference. The summation of contributions from the non-resonant (grey) states leads to the formation of a giant wave, $|\rho_{\neq b}\rangle$, shown in {\bf b}. The excited state, $|b\rangle$ is shown for comparison with a magnification factor of $10$ (dotted, black line in {\bf{b}}). Both wavefunctions are computed for helium using CIS. {\bf c}, Photoelectron spectra from the total analytical model containing contributions from both ground $|a\rangle$ and excited $|b\rangle$ states with resonant atomic excitation: $\Delta\omega=0$. {\bf d}, same as {\bf c}, but with $\Delta\omega=62$ meV. The dashed, white lines denote the expected kinetic energy ($23.3076$ eV) of a photoelectron that has absorbed two resonant photons.}
\label{fig3}
\end{figure*}

We propose that the two-photon signal from $|a\rangle$ can compete with the one-photon signal from $|b\rangle$ due to constructive addition of non-resonant intermediate $p$-states (see grey bound and continuum states in Fig.~\ref{fig3}a). This leads to a giant localized wave:  $|\rho_{\ne b}\rangle$, in comparison with the normalized wavefunction for $|b\rangle$, as shown in Fig.~\ref{fig3}b. The scaling factor due to atomic effects is estimated to be $\sim 1:10^4$ in favor of the two-photon process (see Extended Data Table 1 for the matrix elements). Thus, we can explain why the XUV-FEL pulse is intense enough for the {\it non-resonant} two-photon process from the ground state to be comparable with the one-photon process from the {\it resonant} excited state. 
In general, addition of two pathways leads to quantum interference that depends on the relative phase. From Fig.~\ref{fig3}b we can notice that the giant wave oscillates out of phase with the excited state close to the atomic core, which affects the signs of the matrix elements (see Extended Data Table 1). The ultrafast build-up of the AT doublet can be used to study the resulting interference phenomenon in time. In order to understand this phenomenon, we have used the analytical model to perform calculations where the one- and two-photon contributions are added coherently to simulate the angle-integrated measurements. In Fig.~\ref{fig3}c we show how the resonant case ($\Delta\omega=0$) leads to a strongly asymmetric AT doublet after one Rabi period. Fig.~\ref{fig3}d indicates that a blue detuning ($\Delta\omega=62$ meV) leads to the symmetric AT doublet at an earlier time between $0.5$ and $1$ Rabi period. The advancement of the AT doublet in time follows from the faster Rabi cycling at the rate of generalized Rabi frequency. The required blue shift for the symmetric AT doublet is the signature of an interplay between the one-photon and two-photon processes that depends on the exact pulse form. A general loss of contrast in the AT doublet structures is found by considering the effect of an extended gas target in our model. However, the two-photon doublet was found to be less sensitive to the volume averaging effect when compared to the one-photon doublet (see Methods for details), allowing us to clearly observe the AT doublet in the measured photoelectron signal.

\begin{figure*}[!htb]
\centering
\includegraphics[width=\textwidth]{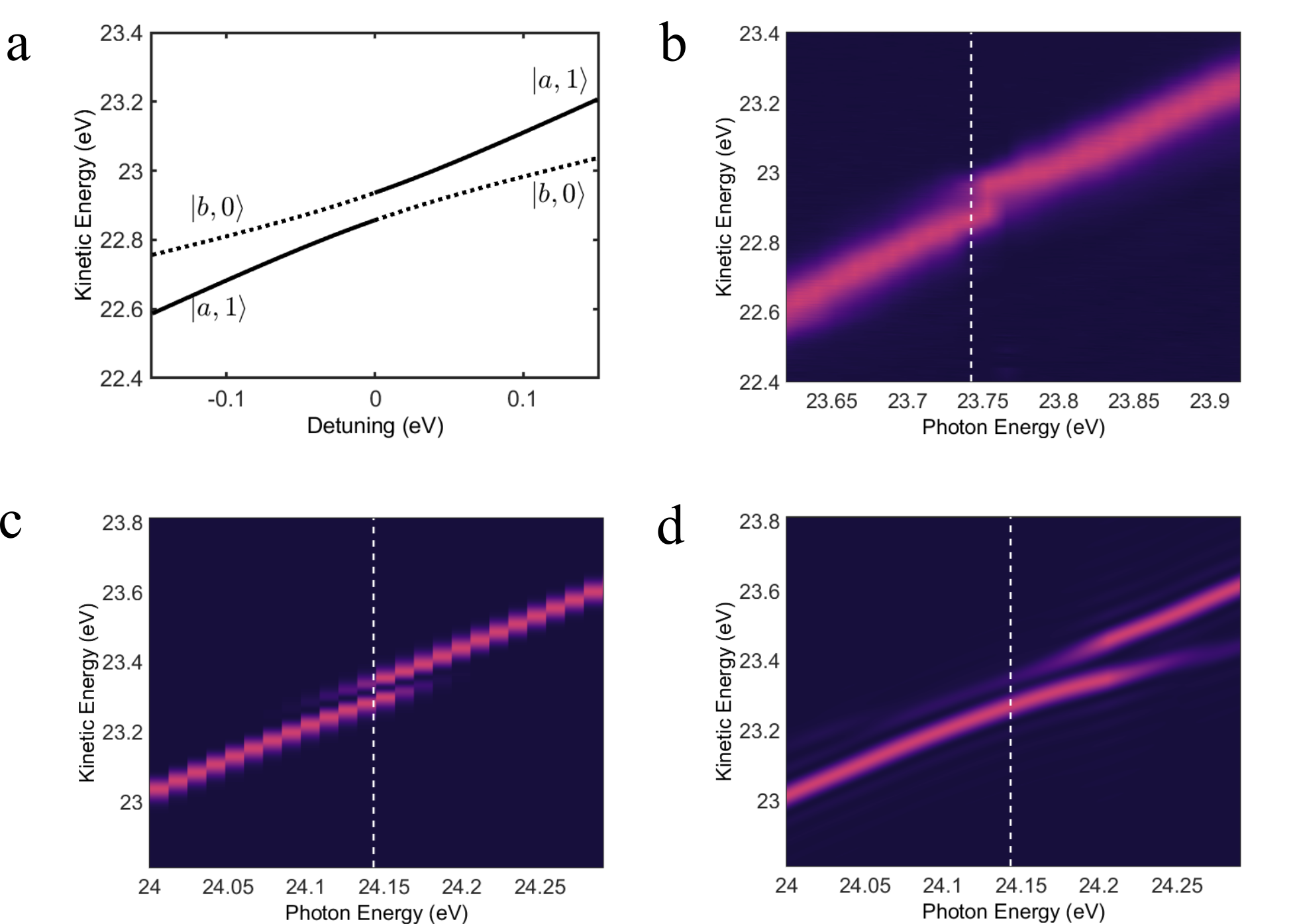}
\caption{{\bf Avoided crossing phenomena in the energy domain.} {\bf a}, Photoelectron kinetic energies, for one photon above the energy of the dressed-atom states, as a function of detuning. Photoelectron spectra as a function of the photon energy retrieved {\bf b}, experimentally, {\bf c}, using TDCIS, and {\bf d}, using the total analytical model for $3/2$ Rabi periods. In each case, the dashed white line corresponds to the photon energy for the $1s^2 \rightarrow 1s4p$ transition in helium. The shifts in energy scales between {\bf a-b}, and {\bf c-d}, are due to the difference between experimental and Hartree-Fock ionization potential. }
\label{fig4}
\end{figure*}

To provide further evidence in support of the coherent interaction between the helium atoms and the XUV-FEL pulses, we show that the ultrafast emergence of the AT doublet can be interpreted in terms of the dressed-atom picture with coupled atom--field energies: $\epsilon_\pm=(\epsilon_a+\epsilon_b+\hbar \omega \pm \hbar W)/2$. One photon energy above these coupled energies implies final photoelectron kinetic energies: $\epsilon^\mathrm{kin}_\pm=\epsilon_\pm+\hbar\omega-I_p$, where $I_p=-\epsilon_a=24.5873$ eV is the ionization potential of helium.
In Fig.~\ref{fig4}a, kinetic energies are labeled with the uncoupled atom--field states\cite{tannoudji2004}, $|a,1\rangle$ and $|b,0\rangle$. The experimental results in Fig.~\ref{fig4}b can be understood as one photon above $|a,1\rangle$ at large detuning of the XUV-FEL pulse. This is because the interaction there is weak and the atom remains mostly in its ground state, $|a\rangle$, such that two photons are required for ionization. In contrast, both coupled energies appear close to the resonance to form an {\it avoided crossing} in kinetic energy. This is a clear signature of coherent Rabi dynamics taking place in the two-level atom. Note that the avoided crossing appears at a blue detuning from the resonant transition, $\Delta\omega=0$ (denoted by the dashed vertical line), revealing the quantum interference between the two pathways from the ground state $|a\rangle$ and the excited state $|b\rangle$. Similar results were obtained from the TDCIS simulations (Fig.~\ref{fig4}c) and the analytical model with contributions from both $|a\rangle$ and $|b\rangle$ for 3/2 Rabi periods (Fig.~\ref{fig4}d). The observed blue detuning for the experimental avoided crossing ($\sim 11$ meV) is very well reproduced by TDCIS calculations ($\sim 14$ meV). The enhanced shift of the AT doublet to blue detuning in the analytical model is an effect of the pulse-envelope that can be reproduced with TDCIS using smoothed flat-top pulses. 

Our results show that it is now possible to simultaneously drive and interrogate ultrafast coherent processes using XUV-FEL pulses. Prior attempts to understand Rabi dynamics at short wavelengths have relied on the strong-field approximation, where the influence of the atomic potential is neglected, leading to an inconsistent AT doublet when compared with numerical simulations\cite{lagattuta1993,girju2007}. In contrast, our analytical model includes the full effect of the atomic potential and Rabi dynamics in the two-level subspace, while the remaining transitions to and within the complement of the Hilbert space are treated by time-dependent perturbation theory. Consequently, we could establish a unique mechanism in the form of a giant Coulomb-induced wave from the ground state to explain why the non-resonant two-photon process can compete with the resonant one-photon process and generate quantum interference effects at the high intensities provided by the XUV-FEL beam. With this model, we now understand how Rabi oscillations can prevail at short wavelengths despite photoionization losses from the neutral atom. Together with our experimental approach of using two-photon ionization as an in-situ probe of the coherent population transfer, which does not rely on any additional laser probe field, the scheme proposed here becomes applicable to other quantum systems as well. With the ongoing developments of seeded FEL facilities around the world\cite{nam2021,liu2022} capable of providing coherent light pulses down to few-\r{a}ngstrom wavelength, our findings can inspire future studies involving core-level electrons in multi-electronic targets, such as molecules, and nano-objects at ultrafast timescales.

\clearpage

\begin{methods}
\section*{Experiment}
The experiment was carried out at the Low Density Matter (LDM) beamline of FERMI\cite{svetina2015}. A pulsed Even-Lavie valve, synchronized with the arrival of the FEL pulse served as the target source. The target gas jet was estimated to be a cone with $2$ mm diameter at the interaction region. We measured the photoelectron spectra at and around the $1s^2 \rightarrow 1s4p$ transition in helium, using a $2$-m long magnetic bottle electron spectrometer (MBES). The gas jet, FEL beam, and magnetic bottle axes are mutually perpendicular, with the first two being on the horizontal plane of the laboratory, and the last one in the vertical direction. Before entering into the flight tube of the MBES, the photoelectrons were strongly retarded to below $1$ eV kinetic energy to achieve high spectral resolution ($E/\Delta E \approx 50$). To suppress any short-term fluctuation arising from the instability of the FEL, we performed a ‘round-trip’ scan across the wavelength range, $52.50 \leftrightarrow 51.80$ nm. Empirically, the FWHM of the XUV-FEL pulse duration ($\tau_{\mathrm{xuv}}$) can be approximated\cite{finetti2017} to be in between $\left(\tau_{\mathrm{seed}}/n^{1/2}\right)$ and $\left(7\tau_{\mathrm{seed}}/6n^{1/3}\right)$. 
Here, $\tau_{\mathrm{seed}}\approx 100$ fs is the duration (FWHM) of the seed pulse (wavelength: $261.08$ nm) and $n=5$ is the harmonic order for the undulator. It leads to $\tau_{\mathrm{xuv}}=56\pm 13$ fs that matches very well the FWHM of $\sim 66$ fs, obtained from the simulation of the FEL dynamics using PERSEO\cite{giannessi2006}. The spectral bandwidth (FWHM) of the pulse was estimated using PERSEO to be around $0.13$ nm at the central wavelength of $\lambda=52.216$ nm. Extended Data Fig. 1a and 1b displays the simulated temporal and spectral profiles of the FEL pulse, respectively. 
At best focus, the spot-size (FWHM) was estimated to be $12$ $\mu$m. We measured the energy per pulse at the output of the FEL undulator to be around $87$ $\mu$J, which refers to the full beam including all photons contained in the transverse Gaussian distribution. In order to consider those, we used $4\sigma$ as the FEL beam diameter at best focus, where $\sigma = 12/2.355 \approx 5.1$ $\mu$m. Hence, the beam waist ($w_0$) is given by $w_0=2\sigma=10.2$ $\mu$m, along with a Rayleigh length of: $\pi w_0^2/\lambda\approx 6.3$ mm. 
\section*{Data analysis}
To filter the measured photoelectron spectra on a shot-to-shot basis, we used the photon spectrum recorded by the Photon Analysis Delivery and REduction System (PADRES) at FERMI to determine the bandwidth (FWHM) of the XUV pulse. Any shot without the photon spectrum was rejected: out of $355000$ shots, $354328$ shots were retained. All the shots having more than $65$ meV FWHM width were discarded (see, Extended Data Fig. 2a). Note that the simulated value of the photon bandwidth ($59$ meV) lies within the filtering window of $20$ – $65$ meV. Additionally, we chose only the shots with integrated spectral intensities ranging from $0.8\times 10^5$ to $1.6\times 10^5$ in arbitrary units (see, Extended Data Fig. 2b). The filtered shots were sorted into $30$ photon-energy-bins, uniformly separated from each other by $\sim 13$ meV and covering the entire photon energy window of the wavelength scan (see, Extended Data Fig. 2c). Overall, only $304192$ shots (filtering ratio of $0.857$) out of the raw data were retained. The measured photoelectron spectra, following shot-to-shot filtering, are shown in Extended Data Fig. 3. The avoided crossing is only faintly visible here. To obtain the clear avoided crossing from Fig. 4b, we deconvoluted the photoelectron spectra for three photon energies near the $1s^2 \rightarrow 1s4p$ transition using the Richardson-Lucy blind iterative algorithm\cite{fish1995}. To reduce the noise introduced during the deconvolution we incorporated the Tikhonov-Miller regularization procedure into the algorithm\cite{dey2004}. The outcomes are shown in Extended Data Fig. 4. Following deconvolution, the values of FWHM for the Gaussian instrument-response-functions were found to be: $70.9\pm 1.2$ meV, $69.6\pm 2.4$ meV and $69.4\pm 1.4$ meV, for the three photon energies. These values match well the combined resolution of $\sim 65$ meV, obtained from the photon bandwidth and the kinetic energy resolution of the MBES. No filter, either metallic or gaseous, was used along the path of the FEL beam. Hence, a minor contribution ($< 5\%$) from the second-order light can be noticed as an asymmetric tail close to $22.8$ eV kinetic energy (see, Extended Data Fig. 4a and 4b). To rule out any artifact from the fluctuations of the FEL pulse properties, we used another filtering criterion for the photon bandwidth ($0$ - $45$ meV) and the integrated spectral intensity ($1\times 10^5$ - $3\times 10^5$ arb. u.). The corresponding deconvoluted photoelectron spectra at $23.753$ eV is shown in Extended Data Fig. 5, along with that from Fig. 2a of the main text. No significant change in the AT doublet structure due to change in filtering criteria could be seen. Finally, for a transform-limited Gaussian pulse, $\tau_{\mathrm{xuv}}$ can vary between $30$ - $90$ fs from shot to shot that encompasses its empirical value: $56\pm13$ fs. Since $\tau^2_{\mathrm{xuv}}$ is significantly higher than the absolute value of the simulated group-delay-dispersion of the FEL pulse: $-690$ fs$^2$, no effect due to the linear chirp was considered in the theoretical calculations.
\section*{Intensity averaging over macroscopic interaction volume}
To better understand the experimental results, we perform an intensity averaging procedure of the analytical model results (see Supplementary Information for details about the model). We assume a Gaussian beam profile with intensity varying as
\begin{equation}\label{eq:intensity}
    I(\rho,z) = I_0 \frac{w_0^2}{w(z)^2}\exp\left[-\frac{2\rho^2}{w(z)^2}\right],
\end{equation}
with $\rho^2 = x^2 +y^2$, $w_0$ the beam waist at focus, and $w(z) = w_0\sqrt{1+(z/z_R)^2}$, where $z_R$ is the Rayleigh length of the beam. We estimate that $w_0 \approx 10 $ $\mu$m and $z_R \approx  6.3$ mm (see, Experiment). Since the target gas is spread out over a finite volume, the experimental signal will contain the response of atoms subject to a range of different intensities according to Eq.~\ref{eq:intensity}. The total signal strength is given by
\begin{equation}
    S(\epsilon) = \int_{V_{\mathrm{gas}}}dV|c(\epsilon,I)|^2,
\end{equation}
where $c(\epsilon,I)$ is the spectral amplitude for energy $\epsilon$ calculated with the model at intensity $I$, and $V_{\mathrm{gas}}$ represents the extent of the gas target. Since the beam waist is much smaller than the extent of the target in the transverse directions, we treat the target as a box shape with an extent of $L = 2$ mm along the $z$-axis. In the transverse direction we consider contributions that are closer than $5w_0$ from the axis of the beam. The total signal is then
\begin{equation}
    S(\epsilon) = 2\pi\int_{-L/2}^{+L/2}dz\int_{0}^{5w_0}d\rho\rho|c(\epsilon,I)|^2.
\end{equation}
The integrals are evaluated using numerical routines from the SciPy library\cite{2020SciPy-NMeth}.

Extended Data Fig. 6a contains a comparison of the shape of the intensity averaged spectrum and the spectrum for a single atom subject to the peak intensity, at the detuning where the peaks appear symmetric ($\Delta \omega = 62$ meV). Extended Data Fig. 6b and 6c contains the same type of curves for the separate contributions from the two- and one-photon process, but at zero detuning (where they are symmetric), illustrating that the two processes are affected differently by the intensity averaging procedure. The spectra in Extended Data Fig. 6b and 6c are normalized to the maximum of each simulation type (single-atom or macroscopic average).  The shape of the two-photon signal is less affected by intensity averaging than the one-photon signal, since most of the signal comes from areas with high intensity due to the quadratic dependence on the electric field in the amplitude. On the other hand, the one-photon signal should have significant contributions from a larger volume of the target, since the one-photon amplitude scales linearly with the strength of the electric field. This means that the shape of the one-photon signal might become more distorted, as seen in Extended Data Fig. 6c, but that its relative contribution compared to the two-photon process should increase, when compared to the single-atom case.
\section*{Numerical simulations using TDCIS}
 The ab initio numerical simulations are performed using the Time-Dependent (TD) Configuration-Interaction Singles (CIS) method\cite{greenman2010,rohringer2006,bertolino2020,bertolino2021} in the velocity gauge. The CIS basis for helium is constructed using Hartree-Fock (HF) orbitals that are computed using B-splines. Exterior complex scaling is used to dampen spurious reflections during time propagation of TDCIS\cite{simon1979}. The vector potential of the XUV-FEL pulse is defined as, 
 \begin{equation}
    A(t) = A_0 \sin(\omega t) \exp\left[ -2 \ln(2) \frac{t^2}{\tau^2} \right].
\end{equation}
 The central frequency, $\omega$, is set close to the CIS atomic transition frequency, $\omega_{ba}=0.887246\,\mathrm{[a.u.]}$ $=24.1432\,\mathrm{eV}$, between the HF ground state: $|a\rangle=1s^2(^1S_0)$ and the singly excited state: $|b\rangle=1s4p(^1P_1)$. The duration of the pulse is set to $\tau= 56\, \mathrm{fs}$ and the peak intensity is set to $I=(\omega A_0)^2\, \mathrm{[a.u.]} \times
3.51\cdot 10^{16}\, \mathrm{[W/cm^2}]=2\times 10^{13}\, \mathrm{W/cm^2}$. The CIS dipole matrix element between the ground state and the excited state is given by $z_{ba} = 0.124a_0$ and the ionization potential is related to the $1s$ orbital energy in HF: $I_p = -\epsilon_{a} = 24.9788\,\mathrm{eV}$, in accordance with Koopmans' theorem. Photoelectron distributions are captured using the time-dependent surface flux (t-SURFF)\cite{tao2012} and infinite-time surface flux (iSURF)\cite{morales2016} methods.
The high kinetic energy of the photoelectrons ensures a proper description of the physics by the surface methods.

\subsection{Data availability} All the relevant data that supports the main findings of the paper are included here. Any additional data related to this study are available from the corresponding authors upon request.

\end{methods}

\begin{addendum}
 \item We acknowledge financial support from LASERLAB-EUROPE (grant agreement no. 654148, European Union’s Horizon 2020 research and innovation programme). S.N. thanks CNRS and F\'{e}d\'{e}ration de Recherche Andr\'{e} Marie Amp\`{e}re, Lyon for financial support. J.M.D. acknowledges support from the Knut and Alice Wallenberg Foundation (2017.0104 and 2019.0154), the Swedish Research Council (2018-03845) and the Olle Engkvist's Foundation (194-0734). R.F. thanks the Swedish Research Council (2018-03731) and the Knut and Alice Wallenberg Foundation (2017.0104) for financial support. P.E.-J. acknowledges support from the Swedish Research Council (2017-04106) and the Swedish Foundation for Strategic Research (FFL12-0101).
 \item[Author contributions] E.O. and M.B. contributed equally to this work. D.B., C.C., M. Di F., P.E.-J., R.F., G.G., M.G., S.M., L.N., J.P., O.P., K.C.P., R.J.S., S.Z. and S.N. performed the experiment. E.O., M.B., S.C., F.Z. and J.M.D. provided the theoretical calculations. E.O. developed the analytical model and M.B. carried out the TDCIS calculations. S.N. analyzed the data. M. Di F., O.P. and C.C. managed the LDM end-station. R.J.S. and R.F. operated the magnetic bottle electron spectrometer. L.B., M.B.D., F.S., and L.G. optimized the machine. M.Ma. and M.Z. characterized the pulses. P.V.D. contributed to the analysis of obtained results. M.Me. and C.M. contributed to the initial planning of the project. S.N. and J.M.D. wrote the manuscript, which all authors discussed. S.N. led the project.
 \item[Competing interests] The authors declare no competing interests.
 \item[Additional information] Correspondence and requests for materials should be addressed to J.M.D. (email: marcus.dahlstrom@matfys.lth.se) or S.N. (email: saikat.nandi@univ-lyon1.fr).
\end{addendum}

\clearpage

\begin{figure*}[!htb]
\centering
\includegraphics[width=\textwidth]{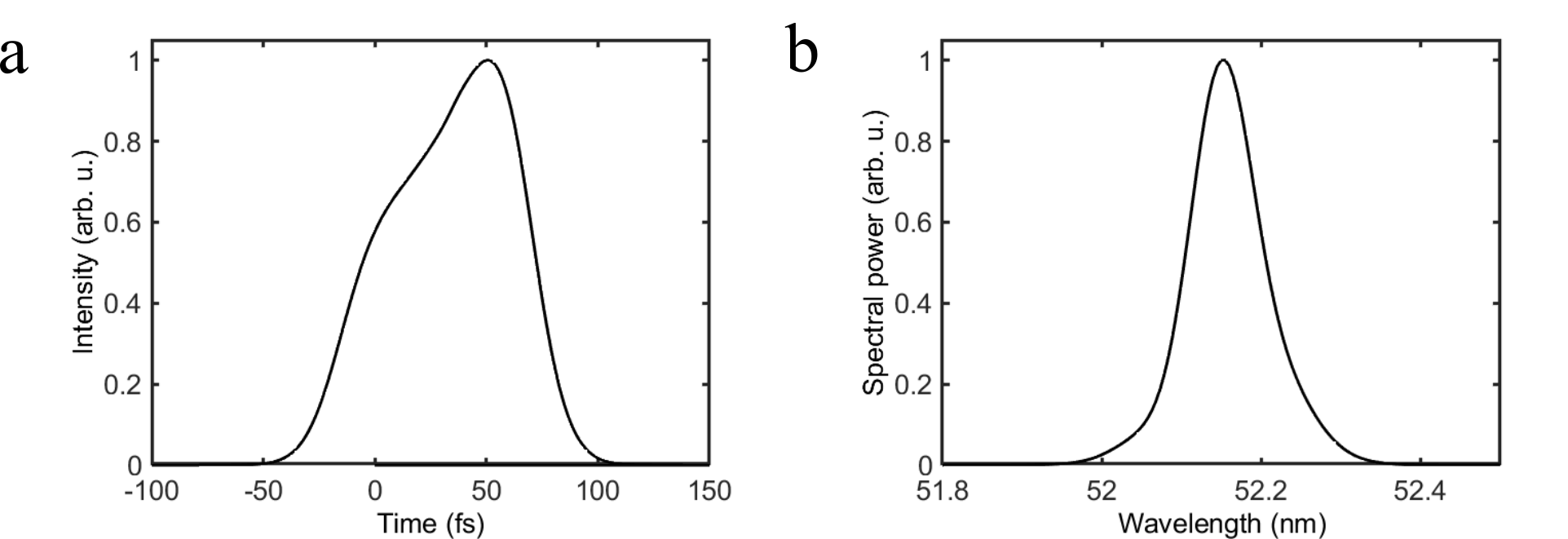}
\caption*{{\bf Extended Data Figure 1. Simulated FEL pulse properties.} {\bf a}, Temporal and {\bf b}, spectral profiles of the XUV-FEL pulse as obtained from the simulations using PERSEO.}  
\end{figure*}

\clearpage

\begin{figure*}[!htb]
\centering
\includegraphics[width=\textwidth]{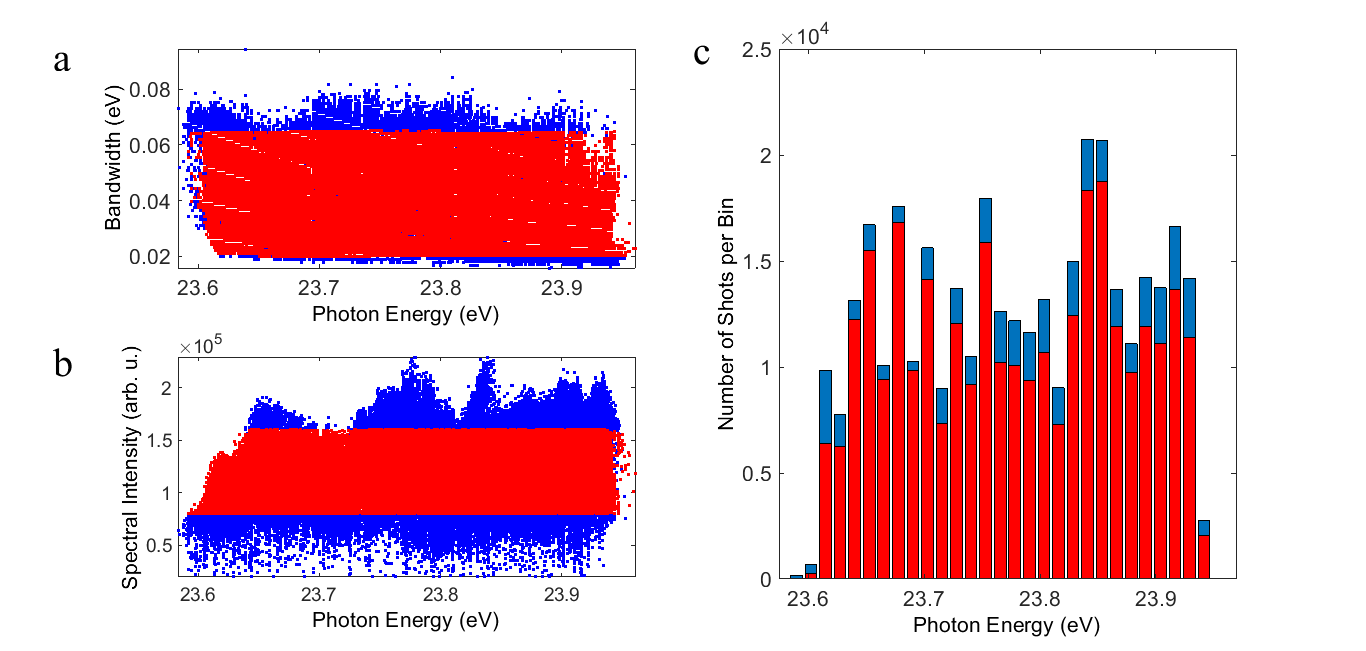}
\caption*{{\bf Extended Data Figure 2. Filtering criteria for the measured data.} {\bf a}, Shot-to-shot variation of the FEL bandwidth (FWHM) as a function of the photon energy. The blue dots represent the measured FWHM using the PADRES spectrometer and the red dots represent the filtered shots. {\bf b}, Same as {\bf a}, but for the integrated spectral intensity. {\bf c}, All the shots are distributed over $30$ equally spaced (spacing: $\sim 13$ meV) photon-energy bins, spanning the entire range of wavelength scan. The red ones correspond to the filtered shots satisfying both the criteria in panel {\bf a}-{\bf b}.}  
\end{figure*}

\clearpage

\begin{figure*}[!htb]
\centering
\includegraphics[width=\textwidth]{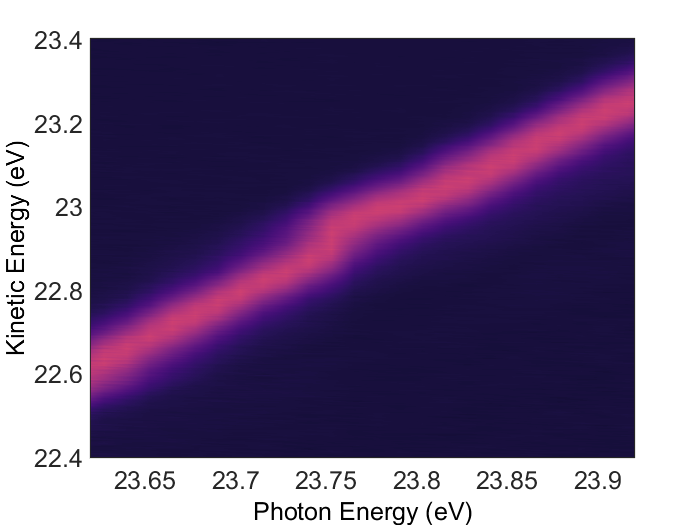}
\caption*{{\bf Extended Data Figure 3. Avoided crossing without deconvolution.} Measured photoelectron spectra, as a function of the photon energy, without any deconvolution procedure performed. Notice the faint avoided crossing.}  
\end{figure*}

\clearpage

\begin{figure*}[!htb]
\centering
\includegraphics[width=\textwidth]{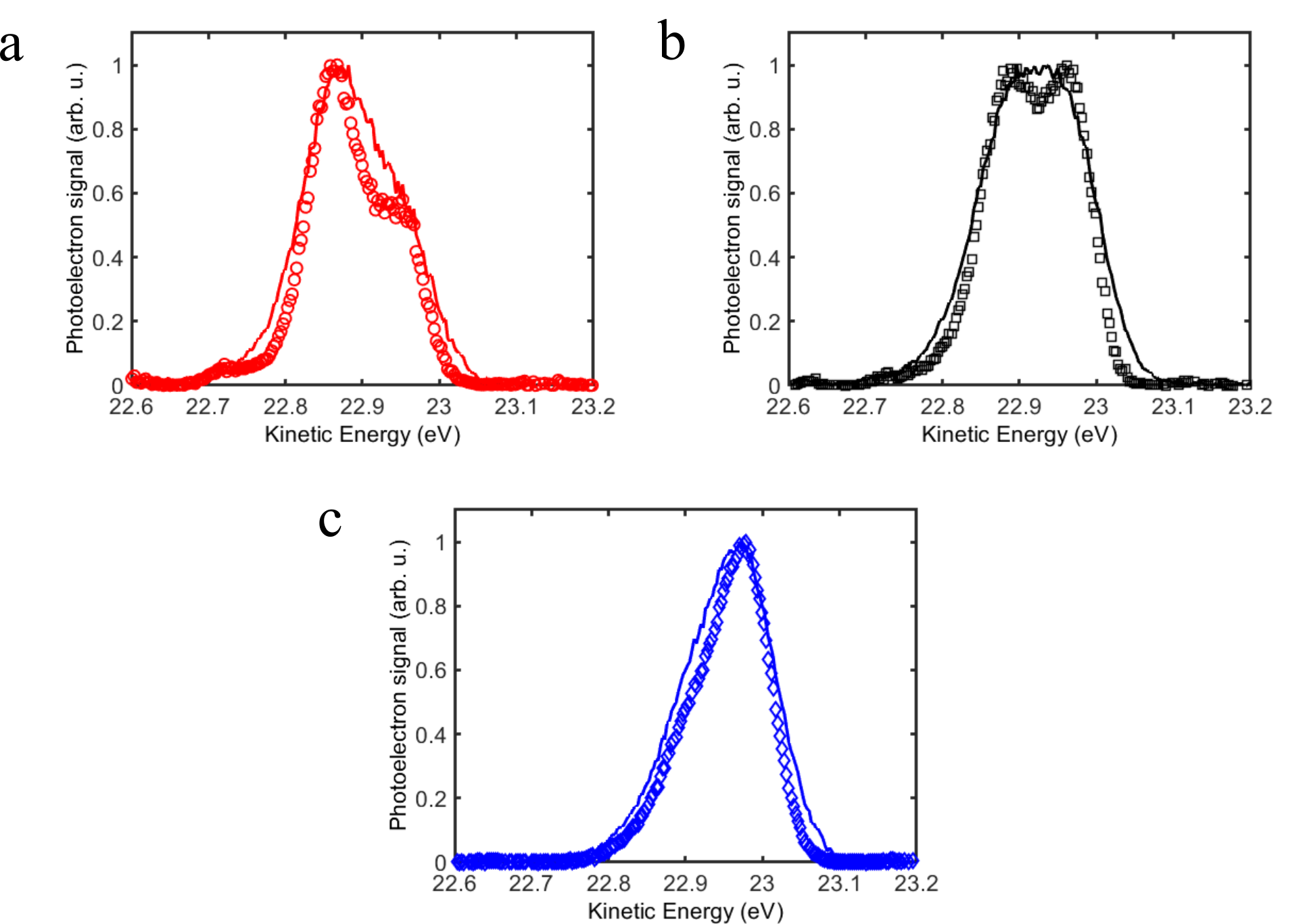}
\caption*{{\bf Extended Data Figure 4. Experimental photoelectron spectra with and without deconvolution.} {\bf a}, Experimental photoelectron spectra at 23.740 eV photon energy. Solid red line: raw data and red circles: deconvoluted form. {\bf b}, Same as {\bf a}, but at $23.753$ eV photon energy. Here, the solid black line represents the raw data and the black squares represents the deconvoluted spectrum. {\bf c}, Same as {\bf a}-{\bf b}, but at $23.766$ eV photon energy, where once again the solid blue line constitutes the raw data and the blue diamonds are for its deconvoluted form. In each case, the open symbols are same as shown in Fig. 2a of the main text.}  
\end{figure*}

\clearpage

\begin{figure*}[!htb]
\centering
\includegraphics[width=\textwidth]{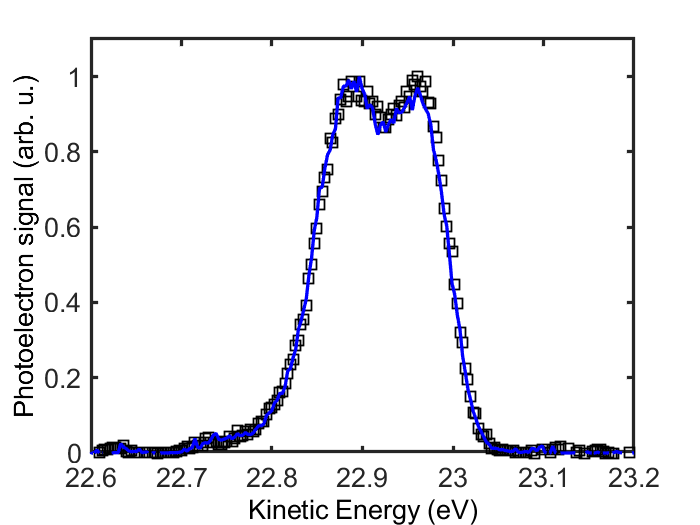}
\caption*{{\bf Extended Data Figure 5. Deconvoluted spectra with two different filtering criteria.} Deconvoluted experimental photoelectron spectra at $23.753$ eV for two different filtering criteria. Open black squares: for photon bandwidth of $20$ – $65$ meV and integrated spectral intensity of $0.8\times 10^5$ - $1.6\times 10^5$ (same as in the Fig. 2a of main text). Solid blue line: for photon bandwidth of $0$ – $45$ meV and integrated spectral intensity of $1\times 10^5$ - $3\times 10^5$.}  
\end{figure*}

\clearpage
\begin{figure*}[!htb]
\centering
\includegraphics[width = \textwidth]{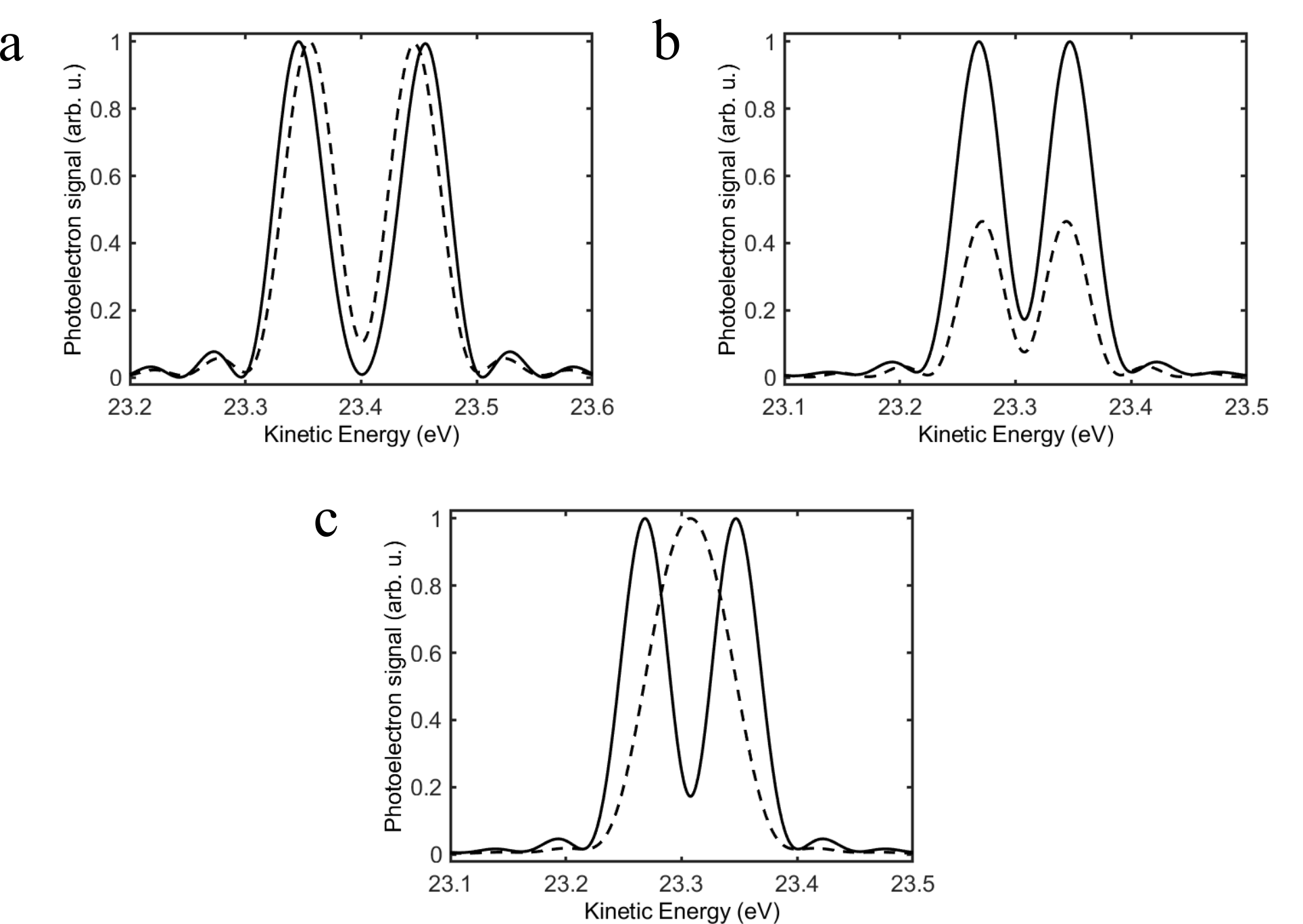}
\caption*{{\bf Extended Data Figure 6. Effects of intensity averaging on the photoelectron spectrum.} {\bf a}, Photoelectron spectra generated with the analytic model for a single atom (solid line), and for a macroscopic sample (dashed line). A pulse length of $3/2$ Rabi periods, and detuning of $\Delta\omega = 62$ meV was used. {\bf b}-{\bf c}, Same as {\bf a}, but for the individual contributions of the two- and one-photon processes, respectively. The results in both {\bf b}, and {\bf c} are calculated for $\Delta\omega = 0$ meV, where the spectra are symmetric. The separate contributions are normalized to the maximum of the one-photon spectra for both the single-atom and intensity averaged signals.}
\end{figure*}

\clearpage
\begin{table}
    \centering
    \caption*{{\bf Extended Data Table 1.} Values of the dipole transition elements $z^\ell_i$ computed using CIS functions.}
    \begin{tabular}{ccc}
         \hline
         \hline 
         Final state & $|b\rangle$ & $|\rho_{\neq b}\rangle$ \\
         \hline
          $s$-wave & 0.009311 &0.1056\\
          $d$-wave & 0.01298 &-1.300 \\
          \hline
          \hline
    \end{tabular}
    \label{tab:dipoles}
\end{table}

\end{document}


\baselineskip24pt

\section*{Supplementary Information} 

\subsection*{Single-atom model}
The Rabi amplitudes for two atomic states: $|a\rangle$ and $|b\rangle$, coupled by an oscillating interaction, are given by \cite{rabi1937,autler1955} (atomic units are used unless otherwise stated):
\begin{align}\tag{S1}
    &a(t)= [\cos(W t/2)-i(\Delta \omega/W)\sin(W t/2)]\exp(i\Delta \omega t /2),
\label{eq:rabia} \\ \tag{S2}
    &b(t)=-i\exp(-i \Delta\omega t/2)\frac{\Omega}{W}
\sin\left(\frac{Wt}{2}\right)
\label{eq:rabib}
\end{align}
where the generalized Rabi frequency is $W = \sqrt{\Omega^2 + \Delta\omega^2}$,
with Rabi frequency, $\Omega$, and detuning, $\Delta\omega=\omega-\omega_{ba}$, with respect to the atomic resonance energy, $\omega_{ba}=\epsilon_b-\epsilon_a$. The Rabi wave packet: $|\Psi_{\mathcal{R}}(t)\rangle=a(t)\exp(-i\epsilon_a t)|a\rangle+b(t)\exp(-i\epsilon_b t)|b\rangle$ is a solution in the two-level subspace  $\mathcal{R}$, spanned by $|a\rangle$ and $|b\rangle$, within the Rotating Wave Approximation (RWA) with the boundary condition $a(0)=1$ and $b(0)=0$. Since the RWA is an excellent approximation for the experimental parameters considered~\cite{autler1955}, $\Omega/\omega_{ba}\approx 0.3\%$, we use it as a zeroth-order solution to our XUV-FEL experiment. In order to obtain corrections, we need to consider the complement to the two-level subspace, which we label: $\mathcal{S}$. These corrections are computed by time-dependent perturbation theory using Dyson equations with coupling to the continuum: $|\epsilon\rangle$ that is part of the $\mathcal{S}$ space and an eigenstate of the atomic Hamiltonian $H_0|\epsilon\rangle = \epsilon|\epsilon\rangle$.  In general the exact propagator: $U$, of the total time-dependent Hamiltonian: $H(t)=H_0+V(t)$, is not known analytically, but using the Rabi amplitudes in Eqs.~(\ref{eq:rabia})-(\ref{eq:rabib}), we can construct an excellent approximation to the exact propagator in the $\mathcal{R}$ space as: 
\begin{align}\tag{S3}
U_{\mathcal{R}}(t,0)|a\rangle = a(t)\exp[-i\epsilon_a(t)]|a\rangle+b(t)\exp[-i\epsilon_b(t)]|b\rangle.
\end{align}
The full propagator can be written as a Dyson equation:
\begin{align}\tag{S4}
    U(t,0)=U_\mathcal{R}(t,0)-i\int _0^t dt' U(t,t')V_{\perp \mathcal{R}}(t')U_{\mathcal{R}}(t',0)
\end{align}
where the interaction has the internal $\mathcal{R}$-space interaction removed: $V_{\perp \mathcal{R}}(t)=V-RV(t)R$, where $R$ is a projector on $\mathcal{R}$. The field-free atomic propagator:  
\begin{align}\tag{S5}
    &U_0(t',t)=\sum _c \! \! \! \! \! \! \! \!  \int U_{c}(t',t) = \sum _c \! \! \! \! \! \! \! \! \int |c\rangle\langle c|\exp[-i\epsilon_c(t'-t)],
\end{align}
does not couple $\mathcal{R}$ and $\mathcal{S}$ because it is diagonal in the basis of atomic eigenstates: $H_0|c\rangle=\epsilon_c|c\rangle$. 
The amplitude for one-photon ionization to the continuum can be approximated to lowest order in $V_{\perp\mathcal{R}}$:
\begin{equation}\tag{S6}
    \alpha_\epsilon(t) =
     -i  \int^{t}_{0}  dt' \langle \epsilon|U(0,t')V_{\perp \mathcal{R}}(t')
     U_{\mathcal{R}}(t',0)|a\rangle
     \approx -i  \int^{t}_{0}  dt' \langle \epsilon | U_\epsilon(0,t')V_{\perp \mathcal{R}}(t') U_b(t',0)|b\rangle b(t'), 
     \label{eq:alpha1}
\end{equation}
where $V_{\perp \mathcal{R}}$ couples from $\mathcal{R}$ to $\mathcal{S}$ within RWA as: 
\begin{align}\tag{S7}
    &\langle \epsilon|V_{\perp \mathcal{R}}(t)|b\rangle = V_{\epsilon b}(t)  = z_{\epsilon b}E(t)\exp[-i\omega t]/2.
\end{align}
This result is valid when two photons are required for photoionization: $\omega \approx \omega_{ba}$ and $2\omega_{ba}>I_p>\omega_{ba}$, where $I_p$ is the binding energy of the atom. In this case, the one-photon ionization process from $\mathcal{R}$ to $\mathcal{S}$ is mediated exclusively from the excited state: $|b\rangle$. The final state, $|\epsilon\rangle$, is propagated with $U_0$, without further interaction with the field in $\mathcal{S}$, as shown on the right-hand side of Eq.~(\ref{eq:alpha1}). 
This first-order amplitude can be evaluated analytically as
\begin{align}\tag{S8}
\label{eq:alpha1full}
\begin{split}
\alpha_{\epsilon}^{(1)}(t) &=-\frac{z_{\epsilon b}E_0\Omega}{2W}\int_0^{t}dt^{\prime}\exp[i(\delta_{\epsilon}-3/2\delta\omega)t^{\prime}]\sin\left(\frac{W t^{\prime}}{2}\right)\\
&=i\frac{z_{\epsilon b}E\Omega}{2W}\Bigg\{\frac{\sin\left[\frac{t}{2}\left(\frac{W}{2} - \frac{3}{2}\Delta\omega + \delta_{\epsilon}\right)\right]\exp\left[i\frac{t}{2}\left(\frac{W}{2} - \frac{3}{2}\Delta\omega +\delta_{\epsilon}\right)\right]}{\frac{W}{2} - \frac{3}{2}\Delta\omega + \delta_{\epsilon}}-\\
&-\frac{\sin\left[\frac{t}{2}\left(\frac{W}{2} + \frac{3}{2}\Delta\omega - \delta_{\epsilon}\right)\right]\exp\left[-i\frac{t}{2}\left(\frac{W}{2} + \frac{3}{2}\Delta\omega - \delta_{\epsilon}\right)\right]}{\frac{W}{2} + \frac{3}{2}\Delta\omega - \delta_{\epsilon}}\Bigg\},
\end{split}
\end{align}
where the relative continuum energy is defined as: 
$\delta_{\epsilon} = \epsilon - \omega_{ba} - \epsilon_b = \epsilon - 2\omega_{ba} -\epsilon_a$ with the photoelectron signal being located at roughly: $\epsilon\approx 2\omega +\epsilon_a$, corresponding to $\epsilon^{\mathrm{kin}} = 2\omega - I_p$.

While the lowest-order expression in Eq.~(\ref{eq:alpha1}) is valid for sufficiently weak XUV-FEL pulses, the present experiment is performed in an intermediate regime where also the second-order interaction with $V_{\perp\mathcal{R}}$ must be investigated. The amplitude for two-photon ionization from state $|a\rangle$, via any accessible intermediate state: $|c\rangle\ne|b\rangle$, can be found in a similar way from the left-hand side of Eq.~(\ref{eq:alpha1}): 
\begin{equation}\tag{S9}
  \alpha_\epsilon^{(2)}(t)=
(-i)^2\int_{0}^t dt'' \int_{0}^{t''} dt' \langle \epsilon|
U_\epsilon (0,t'')V_{\perp R}(t'')U_c(t'',t')V_{\perp R}(t')U_a(t',0)|a\rangle a(t'), 
\end{equation}
where the final and intermediate atomic states are treated to first-order in $V_{\perp R}$ within RWA inside $\mathcal{S}$.
We consider only the final energy region  close to the two photon excitation energy: $\epsilon\approx 2\omega+\epsilon_a$. The second-order contribution in $V_{\perp \mathcal{R}}$ due to two-photon interaction from the excited state: $|b\rangle$, is neglected since it leads to photoelectrons at higher energies: $\epsilon \approx 3\omega+ \epsilon_a$. The amplitude $\alpha_{\epsilon}^{(2)}(t)$ can be written explicitly as,
\begin{align}\tag{S10}
\begin{split}
\alpha_\epsilon^{(2)}(t) &= (-i)^2 \frac{z_{\epsilon c}z_{ca}E_0^2}{8}\int_{0}^t dt'' \int_{0}^{t''} dt' \exp \left[-i(\epsilon_c + \omega - \epsilon)t''\right]\exp\left[ -i(\epsilon_a+\omega - \Delta\omega/2-\epsilon_c)t'\right]\\
&\times\left[\exp[iWt'/2](1-\Delta\omega/W) + \exp[-iWt'/2](1+\Delta\omega/W) \right].
\end{split}
\end{align}
Performing the double integral results in
\begin{align}\tag{S11}
\begin{split}
\alpha_\epsilon^{(2)}(t) &=-i \frac{z_{\epsilon c}z_{ca}E_0^2}{4}\bigg\{\frac{(1-\Delta\omega/W)\exp\left[ i( \frac{W}{2} + \delta_{\epsilon} -\frac{3}{2}\Delta\omega)\frac{t}{2}\right]\sin\left[ (\frac{W}{2} + \delta_{\epsilon} -\frac{3}{2}\Delta\omega)\frac{t}{2}\right]}{\tilde{\omega}^{-}(W/2 + \delta_{\epsilon} -3\Delta\omega/2)}\\
&+\frac{(1+\Delta\omega/W)\exp\left[ -i( \frac{W}{2} - \delta_{\epsilon} +\frac{3}{2}\Delta\omega)\frac{t}{2}\right]\sin\left[ ( \frac{W}{2} - \delta_{\epsilon} +\frac{3}{2}\Delta\omega)\frac{t}{2}\right]}{\tilde{\omega}^{+}( W/2 - \delta_{\epsilon} +3\Delta\omega/2)}\\
&-\frac{(1-\Delta\omega/W)\tilde{\omega}^{+} + (1+\Delta\omega/W)\tilde{\omega}^{-}}{\tilde{\omega}^{-}\tilde{\omega}^{+}(\delta_{\epsilon} +\epsilon_{b}-\epsilon_{c}-\Delta\omega)} \exp\left[i(\delta_{\epsilon} +\epsilon_{b}-\epsilon_{c}-\Delta\omega)\frac{t}{2}\right]\\
&\times\sin\left[(\delta_{\epsilon} +\epsilon_{b}-\epsilon_{c}-\Delta\omega)\frac{t}{2}\right] \bigg\},
\end{split}
\label{eq:alpha2full}
\end{align}
where the quantities $\tilde{\omega}^{\pm} = \epsilon_a+\omega - \epsilon_c -\Delta\omega/2\pm W/2$ have been introduced.
Modulo prefactors, the amplitude $\alpha_\epsilon^{(2)}(t)$ is similar to $\alpha_\epsilon^{(1)}(t)$ in the sense that both are peaked at $\delta_\epsilon = 3\Delta\omega/2 \pm W/2$. We interpret these two peaks as an Autler-Townes doublet, but note that the peaks have different strengths in general with non-zero detuning. Unlike the first-order AT doublet in Eq.~(\ref{eq:alpha1full}), the second-order AT doublet in Eq.~(\ref{eq:alpha2full}) has the same sign on both peaks. This implies that asymmetric quantum interference effects should be expected when photoelectrons are ejected to the same final state with comparable amplitudes in both processes. In addition,  $\alpha_\epsilon^{(2)}(t)$ has a peak at $\delta_\epsilon = \epsilon_c-\epsilon_b+\Delta\omega$ that depends on the intermediate state energy, $\epsilon_c$. The closest state to $1s4p$ is $1s5p$ at $0.3$ eV.

In order to evaluate the second-order amplitude a summation and integration over all accessible intermediate states should be performed. In this process it is plausible that the $\epsilon_c$-dependent additional peaks of $\alpha_\epsilon^{(2)}$ will wash out due to their different energy positions and phases. The summation and integration over all accessible intermediate states implies a perturbed wavefunction (or ``giant wave'' from the main text) of the form: 
\begin{align}\label{eq:rho_not_b}\tag{S12}
    |\rho^\pm_{\ne b}\rangle =\sum_{c\ne b} \! \! \! \! \! \! \! \! \!  \int \frac{|c\rangle z_{ca}}{(\epsilon_a+\omega-\epsilon_c-\Delta\omega/2\pm W/2)},
\end{align}
which can be identified as part of the first and second term in Eq.~(\ref{eq:alpha2full}). Compared to treating the ionization step using the strong field approximation \cite{girju2007}, the method presented here includes the effect of intermediate bound states, but does not include the interaction with the XUV-FEL to all orders for states in $\mathcal{S}$.

For the model results that are presented in the main article, the perturbed wavefunction was constructed using configuration interaction singles (CIS) basis states for He. When computing the perturbed wavefunction, $\Delta\omega$ and $W$ were neglected in the denominator of Eq.~(\ref{eq:rho_not_b}), as they were found to not have a significant impact on the resulting photoelectron spectra. We denote the perturbed state constructed with these approximations as $|\rho_{\neq b}\rangle$. As shown in Fig.~3(a) of the main article, the radial part of the perturbed wavefunction is significantly larger in magnitude when compared to the radial wavefunction of state $|b\rangle$. The reason for the perturbed wavefunction to be so large is that all intermediate states are similar to each other close to the Coulomb potential. This leads to constructive interference and to a large dipole transition element from the intermediate states to the final state as compared to the transition from the $|b\rangle=1s4p$ state to the final state. Only two intermediate states ($1s2p$ and $1s3p$) add destructively to the perturbed wavefunction since they have less energy than the $1s4p$ state  (as shown in Fig. 3(a) of the main text). 

To compute the dipole transition elements to the continuum, CIS continuum functions were used, and the estimated values for a photoelectron kinetic energy of $23.3076$ eV can be found in Extended Data Table~1. Combining the factor from one extra interaction with the electric field, $E_0/2$, with the ratio of the dipole elements it is possible to estimate the ratio of the amplitudes for the one- and two-photon processes
\begin{equation}\label{eq:ratio}\tag{S13}
    R^\ell = \left|\frac{E_0}{2}\frac{z^\ell_{\rho_{\neq b}}}{z^\ell_b}\right|,
\end{equation} where $z^\ell_i$ is the transition matrix element from state $i$ to a final continuum state with orbital angular momentum $\ell$. At an intensity of $2\times10^{13}$~W/cm$^2$, which corresponds to an electric field amplitude: $E_0=\omega A_0=0.023880\, \mathrm{a.u.}$, we estimate $R^s = 0.13546$ and $R^d = 1.1958$. While the one-photon transition is clearly dominant to the $s$-wave, the two contributions to the $d$-wave are more comparable in magnitude. The total signal mainly comes from the two-photon transition to the $d$-wave, but interference with the one-photon transition to the $d$-wave can not be neglected since it will lead to a blue shift of the symmetric AT doublet. In the main article, the photoelectron spectra shown for the model contain the combined signal for both $s$ and $d$ final states. All atomic parameters required for the model are taken from the ab initio TDCIS method described in Methods.